\documentclass[twocolumn,superscriptaddress,showpacs,preprintnumbers,amsmath,amssymb]{revtex4}

\usepackage{graphicx}
\usepackage{dcolumn}
\usepackage{bm}
\usepackage{xspace} 
\usepackage{amsmath}
\usepackage{amssymb}
\usepackage{color}

\def\to {\rightarrow}
\newcommand{\lw}[1]{\smash{\lower1.7ex\hbox{#1}}}
\newcommand{\lww}[1]{\smash{\lower6.7ex\hbox{#1}}}

\def\bellelogo{\vbox to 16mm{
               \vss\hbox to \textwidth{\resizebox{!}{2cm}{
               \includegraphics{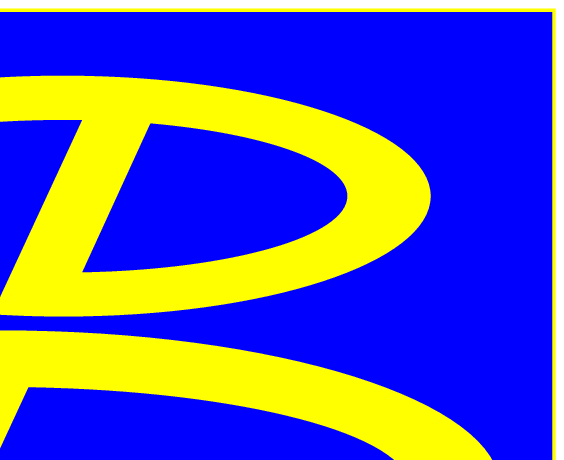}}\hss }}\vspace{-1cm}}

\def\preprintB{\hbox{\hfil Belle Preprint 2006-11}}

\begin{document}

\title{
  \vbox{\bellelogo \vspace{-1cm}
  \hbox to \textwidth{\rm\normalsize\hss\preprintB}
  }
\vspace*{20mm}
  Evidence of the Purely Leptonic Decay $B^{-}\rightarrow\tau^{-}\bar{\nu}_{\tau}$ }

\begin{abstract}
We present the first evidence of the decay $B^{-}\rightarrow\tau^{-}\bar{\nu}_{\tau}$, 
using $414~\textrm{fb}^{-1}$ of data collected at the $\Upsilon(4S)$ resonance 
with the Belle detector at the KEKB asymmetric-energy $e^{+}e^{-}$ collider. 
Events are tagged by fully reconstructing one of the $B$ mesons in hadronic modes.
We detect the signal with a significance of $3.5$ standard deviations 
including systematics, and measure the branching fraction to be 
${\cal B}(B^{-}\rightarrow\tau^{-}\bar{\nu}_{\tau}) = 
 (1.79^{+0.56}_{-0.49}(\mbox{stat})^{+0.46}_{-0.51}(\mbox{syst})) \times 10^{-4}$.
This implies that 
$f_{B} =  0.229^{+0.036}_{-0.031}(\mbox{stat})^{+0.034}_{-0.037}(\mbox{syst})$ GeV
and is the first direct measurement of this quantity.
\end{abstract}

\affiliation{Budker Institute of Nuclear Physics, Novosibirsk}
\affiliation{Chiba University, Chiba}
\affiliation{Chonnam National University, Kwangju}
\affiliation{University of Cincinnati, Cincinnati, Ohio 45221}
\affiliation{University of Frankfurt, Frankfurt}
\affiliation{Gyeongsang National University, Chinju}
\affiliation{University of Hawaii, Honolulu, Hawaii 96822}
\affiliation{High Energy Accelerator Research Organization (KEK), Tsukuba}
\affiliation{Hiroshima Institute of Technology, Hiroshima}
\affiliation{University of Illinois at Urbana-Champaign, Urbana, Illinois 61801}
\affiliation{Institute of High Energy Physics, Chinese Academy of Sciences, Beijing}
\affiliation{Institute of High Energy Physics, Vienna}
\affiliation{Institute of High Energy Physics, Protvino}
\affiliation{Institute for Theoretical and Experimental Physics, Moscow}
\affiliation{J. Stefan Institute, Ljubljana}
\affiliation{Kanagawa University, Yokohama}
\affiliation{Korea University, Seoul}
\affiliation{Kyoto University, Kyoto}
\affiliation{Kyungpook National University, Taegu}
\affiliation{Swiss Federal Institute of Technology of Lausanne, EPFL, Lausanne}
\affiliation{University of Ljubljana, Ljubljana}
\affiliation{University of Maribor, Maribor}
\affiliation{University of Melbourne, Victoria}
\affiliation{Nagoya University, Nagoya}
\affiliation{Nara Women's University, Nara}
\affiliation{National Central University, Chung-li}
\affiliation{National United University, Miao Li}
\affiliation{Department of Physics, National Taiwan University, Taipei}
\affiliation{H. Niewodniczanski Institute of Nuclear Physics, Krakow}
\affiliation{Nippon Dental University, Niigata}
\affiliation{Niigata University, Niigata}
\affiliation{Nova Gorica Polytechnic, Nova Gorica}
\affiliation{Osaka City University, Osaka}
\affiliation{Osaka University, Osaka}
\affiliation{Panjab University, Chandigarh}
\affiliation{Peking University, Beijing}
\affiliation{University of Pittsburgh, Pittsburgh, Pennsylvania 15260}
\affiliation{Princeton University, Princeton, New Jersey 08544}
\affiliation{RIKEN BNL Research Center, Upton, New York 11973}
\affiliation{Saga University, Saga}
\affiliation{University of Science and Technology of China, Hefei}
\affiliation{Seoul National University, Seoul}
\affiliation{Shinshu University, Nagano}
\affiliation{Sungkyunkwan University, Suwon}
\affiliation{University of Sydney, Sydney NSW}
\affiliation{Tata Institute of Fundamental Research, Bombay}
\affiliation{Toho University, Funabashi}
\affiliation{Tohoku Gakuin University, Tagajo}
\affiliation{Tohoku University, Sendai}
\affiliation{Department of Physics, University of Tokyo, Tokyo}
\affiliation{Tokyo Institute of Technology, Tokyo}
\affiliation{Tokyo Metropolitan University, Tokyo}
\affiliation{Tokyo University of Agriculture and Technology, Tokyo}
\affiliation{Toyama National College of Maritime Technology, Toyama}
\affiliation{University of Tsukuba, Tsukuba}
\affiliation{Virginia Polytechnic Institute and State University, Blacksburg, Virginia 24061}
\affiliation{Yonsei University, Seoul}
  \author{K.~Ikado}\affiliation{Nagoya University, Nagoya} 
  \author{K.~Abe}\affiliation{High Energy Accelerator Research Organization (KEK), Tsukuba} 
  \author{K.~Abe}\affiliation{Tohoku Gakuin University, Tagajo} 
  \author{I.~Adachi}\affiliation{High Energy Accelerator Research Organization (KEK), Tsukuba} 
  \author{H.~Aihara}\affiliation{Department of Physics, University of Tokyo, Tokyo} 
  \author{K.~Akai}\affiliation{High Energy Accelerator Research Organization (KEK), Tsukuba} 
  \author{M.~Akemoto}\affiliation{High Energy Accelerator Research Organization (KEK), Tsukuba} 
  \author{D.~Anipko}\affiliation{Budker Institute of Nuclear Physics, Novosibirsk} 
  \author{K.~Arinstein}\affiliation{Budker Institute of Nuclear Physics, Novosibirsk} 
  \author{V.~Aulchenko}\affiliation{Budker Institute of Nuclear Physics, Novosibirsk} 
% ...
\collaboration{The Belle Collaboration}

\maketitle

\begin{figure}
\begin{center}
 \includegraphics[width=.35\textwidth]{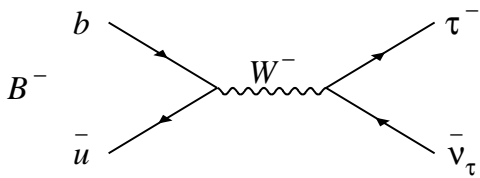}
\caption{Purely leptonic $B$ decay proceeds via quark annihilation into a $W$ boson.}
    \label{fig:decay_Btaunu}
\end{center}   
\end{figure}

In the Standard Model (SM), the purely leptonic decay 
$B^{-}\rightarrow\tau^{-}\bar{\nu}_{\tau}$~\cite{conjugate} proceeds via 
annihilation of $b$ and $\overline{u}$ quarks to a $W^-$ boson (Fig.~\ref{fig:decay_Btaunu}). 
It provides a direct determination of the product of the $B$ meson decay 
constant $f_B$ and the magnitude of the
Cabibbo-Kobayashi-Maskawa (CKM) matrix element $|V_{ub}|$.
The branching fraction is given by
\begin{equation}
 \label{eq:BR_B_taunu}
{\cal B}(B^{-}\rightarrow\tau^{-}\bar{\nu}_{\tau}) 
= \frac{G_{F}^{2}m_{B}m_{\tau}^{2}}{8\pi}\left(1-\frac{m_{\tau}^{2}}
{m_{B}^{2}}\right)^{2}f_{B}^{2}|V_{ub}|^{2}\tau_{B},
\end{equation}
where $G_F$ is the Fermi coupling constant, 
$m_{B}$ and $m_{\tau}$ are the $B$ and $\tau$ masses, respectively, 
and $\tau_B$ is the $B^-$ lifetime~\cite{Eidelman:2004wy}.
The expected branching fraction is  $(1.59\pm 0.40) \times 10^{-4}$
using $|V_{ub}| = (4.39 \pm 0.33) \times 10^{-3}$, determined by inclusive
charmless semileptonic $B$ decay data~\cite{HFAG}, $\tau_{B} = 1.643\pm 0.010$ ps~\cite{HFAG},
and $f_B = 0.216\pm 0.022$ GeV
obtained from lattice QCD calculations~\cite{Gray:2005ad}.
Physics beyond the SM, such as supersymmetry or two-Higgs doublet models,
could modify ${\cal B}(B^{-}\rightarrow\tau^{-}\bar{\nu}_{\tau})$ through
the introduction of a charged Higgs boson~\cite{Hou:1992sy}.
Purely leptonic $B$ decays have not been observed before.
The most stringent upper limit on $B^{-}\rightarrow\tau^{-}\bar{\nu}_{\tau}$
comes from the BaBar experiment: 
${\cal B}(B^{-}\rightarrow\tau^{-}\bar{\nu}_{\tau}) < 2.6 \times 10^{-4}$
(90\% C.L.)~\cite{Aubert:2005}.
In this paper, we present the first evidence for $B^{-}\rightarrow\tau^{-}\bar{\nu}_{\tau}$ from the Belle experiment.

We use a $414~\textrm{fb}^{-1}$ data sample containing 
$449\times 10^{6}$ $B$ meson pairs collected with the Belle detector
at the KEKB asymmetric-energy $e^{+}e^{-}$ ($3.5$ on $8$ GeV) collider~\cite{Kurokawa:2003} 
operating at the $\Upsilon(4S)$ resonance ($\sqrt{s} = 10.58$ GeV). 
The Belle detector~\cite{belle_detector:2003} is a large-solid-angle
magnetic spectrometer consisting of a silicon vertex detector,
a $50$-layer central drift chamber (CDC), a system of aerogel threshold
Cherenkov counters (ACC), time-of-flight scintillation
counters (TOF), and an electromagnetic calorimeter comprised of
CsI(Tl) crystals (ECL)
located inside a superconducting solenoid coil that provides a $1.5$ T
magnetic field. An iron flux-return located outside of the coil is
instrumented to identify $K_{L}^{0}$ and muons.

We use a detailed Monte Carlo (MC) simulation based on GEANT~\cite{GEANT} 
to determine the signal selection efficiency and study the background.
In order to reproduce effects of beam background, data taken with random
triggers for each run period are overlaid on simulated events. 
The $B^{-}\rightarrow\tau^{-}\bar{\nu}_{\tau}$ signal decay is generated 
by the EvtGen package~\cite{EvtGen}.
To model the background from $e^+e^- \to B\overline{B}$ and continuum 
$q\overline{q}~(q = u, d, s, c)$ production processes, large 
$B\overline{B}$ and $q\overline{q}$ MC samples 
corresponding to about twice the data sample
are used.
We also use MC samples for rare $B$ decay processes, such as charmless 
hadronic, radiative, electroweak decays and $b \to u$ semileptonic decays.

We fully reconstruct one of the $B$ mesons in 
the event, referred to hereafter as the tag side ($B_{\rm tag}$), 
and compare properties of the remaining particle(s), referred to as the 
signal side ($B_{\rm sig}$), to those expected for signal and background.
The method allows us to suppress strongly the combinatorial background 
from both $B\overline{B}$ and continuum events.
In order to avoid experimental bias, 
the signal region in data is not examined until the event 
selection criteria are finalized.

The $B_{\rm tag}$ candidates are reconstructed in the following decay modes: 
$B^{+} \rightarrow \overline{D}{}^{(*)0} \pi^{+}$, 
$\overline{D}{}^{(*)0}\rho^{+}$, 
$\overline{D}{}^{(*)0}a_{1}^{+}$ 
and $\overline{D}{}^{(*)0}D_{s}^{(*)+}$.
The $\overline{D}{}^{0}$ mesons are reconstructed as 
$\overline{D}{}^{0}\rightarrow K^{+}\pi^{-}$, $K^{+}\pi^{-}\pi^{0}$,
$K^{+}\pi^{-}\pi^{+}\pi^{-}$, $K_{S}^{0}\pi^{0}$, $K_{S}^{0}\pi^{-}\pi^{+}$,
$K_{S}^{0}\pi^{-}\pi^{+}\pi^{0}$ and $K^{-}K^{+}$, and
the $D_{s}^{+}$ mesons are reconstructed as 
$D_{s}^{+}\rightarrow K_{S}^{0}K^{+}$ and $K^{+}K^{-}\pi^{+}$.
The $\overline{D}{}^{*0}$ and $D_{s}^{*+}$ mesons are reconstructed in
$\overline{D}{}^{*0} \to \overline{D}{}^0 \pi^0, \overline{D}{}^0 \gamma$,
and $D_{s}^{*+} \to D_{s}^{+} \gamma$ modes.  
The selection of $B_{\rm tag}$ candidates is based on the 
beam-constrained mass $M_{\rm bc}\equiv\sqrt{E_{\rm beam}^{2} - p_{B}^{2}}$
and the energy difference $\Delta E\equiv E_{B} - E_{\rm beam}$.
Here, $E_{B}$ and $p_{B}$ are the reconstructed energy and momentum
of the $B_{\rm tag}$ candidate in the $e^+e^-$ center-of-mass (CM) system,
and $E_{\rm beam}$ is the beam energy in the CM frame.
The selection criteria for $B_{\rm tag}$ are defined as
$M_{\rm bc}>5.27~\mbox{GeV}/c^{2}$ and $-80~\mbox{MeV} <\Delta E< 60~\mbox{MeV}$.
If an event has multiple $B_{\rm tag}$ candidates, we choose the one having
the smallest $\chi^{2}$ based on deviations from the nominal values of 
$\Delta E$, the $D$ candidate mass, and the $D^{*} - D$ mass difference if
applicable.
By fitting the $M_{\rm bc}$ distribution to the sum of an empirical 
parameterization of the background shape~\cite{Albrecht:1986nr} plus a 
signal shape~\cite{Bloom:1983pc}, we estimate the number of $B_{\rm tag}$'s 
and their purity in the selected region to be $6.80 \times 10^{5}$ and 
$0.55$, respectively. 

In the events where a $B_{\rm tag}$ is reconstructed, we search for decays
of $B_{\rm sig}$ into a $\tau$ and a neutrino. 
Candidate events are required to have one or three charged track(s) on the
signal side with the total charge being opposite to that of $B_{\rm tag}$.
The $\tau$ lepton is identified in the five decay modes,
$\mu^{-}\bar{\nu}_{\mu}\nu_{\tau}$,
$e^{-}\bar{\nu}_{e}\nu_{\tau}$, 
$\pi^{-}\nu_{\tau}$,
$\pi^{-}\pi^{0}\nu_{\tau}$ and 
$\pi^{-}\pi^{+}\pi^{-}\nu_{\tau}$,
which taken together correspond to $81\%$ of all $\tau$ decays~\cite{Eidelman:2004wy}.
The muon, electron and charged pion candidates are selected based on
information from particle identification subsystems.
The leptons are selected with efficiency greater than 90\% 
for both muons and electrons in the CM momentum region above 1.2 GeV/$c$, and 
misidentification rates of less than 0.2\%(1.5\%) for electrons (muons). 
Kaon candidates are rejected for all charged tracks on the signal side.
The $\pi^0$ candidates are reconstructed by requiring the invariant mass of two
$\gamma$'s to satisfy $|M_{\gamma\gamma}-m_{\pi^0}| < 20~\mbox{MeV}/c^{2}$.
For all modes except $\tau^{-}\rightarrow\pi^{-}\pi^{0}\nu_{\tau}$, we reject events with 
$\pi^{0}$ mesons on the signal side.
We place the following requirements on the track momentum in the CM frame,
$p_{\ell} > 0.3~\textrm{GeV}/c$ for $\mu^{-}\bar{\nu}_{\mu}\nu_{\tau}$ and $e^{-}\bar{\nu}_{e}\nu_{\tau}$,
$p_{\pi^{-}} > 1.0~\textrm{GeV}/c$ for $\pi^{-}\nu_{\tau}$,
$p_{\pi^{-}\pi^{0}} > 1.2~\textrm{GeV}/c$ for $\pi^{-}\pi^{0}\nu_{\tau}$ and
$p_{\pi^{-}\pi^{+}\pi^{-}} > 1.8~\textrm{GeV}/c$ for
$\pi^{-}\pi^{+}\pi^{-}\nu_{\tau}$.
We calculate the missing momentum of the event in the CM frame ($p_{\rm miss}$)
from $p_B$ and the momenta of charged tracks and $\pi^0$'s on the signal side.
We require
$p_{\rm miss} > 0.2~\textrm{GeV}/c$ for $\mu^{-}\bar{\nu}_{\mu}\nu_{\tau}$ and 
$e^{-}\bar{\nu}_{e}\nu_{\tau}$,
$p_{\rm miss} > 1.0~\textrm{GeV}/c$ for $\pi^{-}\nu_{\tau}$,
$p_{\rm miss} > 1.2~\textrm{GeV}/c$ for $\pi^{-}\pi^{0}\nu_{\tau}$ and
$p_{\rm miss} > 1.8~\textrm{GeV}/c$ for $\pi^{-}\pi^{+}\pi^{-}\nu_{\tau}$.
In order to suppress background where particles produced along the beam pipe
escape detection, the cosine of the angle of the missing momentum 
($\cos\theta_{\rm miss}^{*}$) is required to satisfy
$-0.86 < \cos\theta_{\rm miss}^{*} < 0.95$
in the CM frame.
We further require the invariant mass of the visible decay products to satisfy
$|M_{\pi\pi}-m_{\rho}| < 0.15~\mbox{GeV}/c^{2}$ and 
$|M_{\pi\pi\pi}-m_{a_{1}^{-}}| < 0.3~\mbox{GeV}/c^{2}$.
All the selection criteria have been optimized to achieve the highest sensitivity in MC.

The most powerful variable for separating signal and background is the 
remaining energy in the ECL, denoted as $E_{\rm ECL}$, which is sum of
the energies of neutral clusters that are not associated with either the 
$B_{\rm tag}$ or the $\pi^{0}$ candidate from the 
$\tau^{-}\rightarrow \pi^{-}\pi^{0}\nu_{\tau}$ decay.
For neutral clusters contributing to $E_{\rm ECL}$, we require a minimum 
energy threshold of 50 MeV for the barrel and 100 (150) MeV for the forward (backward) endcap ECL.
A higher threshold is used for the endcap ECL because the effect of beam
background is more severe.
For signal events, $E_{\rm ECL}$ must be either zero or a small value 
arising from beam background hits, therefore, signal events peak at 
low $E_{\rm ECL}$.
On the other hand, background events are distributed toward higher 
$E_{\rm ECL}$ due to the contribution from additional neutral clusters.

The $E_{\rm ECL}$ signal region is optimized for each $\tau$ decay mode based
on the MC simulation, and is defined by $E_{\rm ECL} < 0.2~\mbox{GeV}$ for the
$\mu^{-}\bar{\nu}_{\mu}\nu_{\tau}$, $e^{-}\bar{\nu}_{e}\nu_{\tau}$ and
$\pi^{-}\nu_{\tau}$ modes, and $E_{\rm ECL} < 0.3~\mbox{GeV}$ for
the $\pi^{-}\pi^{0}\nu_{\tau}$ and $\pi^{-}\pi^{+}\pi^{-}\nu_{\tau}$
modes. 
The $E_{\rm ECL}$ sideband region is defined by $0.4~\mbox{GeV} < E_{\rm ECL} < 1.2$ GeV 
for the $\mu^{-}\bar{\nu}_{\mu}\nu_{\tau}$, $e^{-}\bar{\nu}_{e}\nu_{\tau}$ and
$\pi^{-}\nu_{\tau}$ modes, and by $0.45~\mbox{GeV} < E_{\rm ECL} < 1.2$ GeV for
the $\pi^{-}\pi^{0}\nu_{\tau}$ and $\pi^{-}\pi^{+}\pi^{-}\nu_{\tau}$ modes.
Table~\ref{tab:signal_yields} shows the number of events found in the sideband 
region for data ($N_{\rm side}^{\rm obs}$) and for the background MC simulation 
($N_{\rm side}^{\rm MC}$) scaled to the equivalent integrated luminosity in data.
Their good agreement for each $\tau$ decay mode indicates the validity of the
background MC simulation.
According to the MC simulation, about 95\% (5\%) of the background events come 
from $B\overline{B} (q\overline{q})$ processes. 
Table~\ref{tab:signal_yields} also shows the number of the background MC events 
in the signal region ($N_{\rm sig}^{\rm MC}$).  
The MC simulation predicts that the background in the signal region comes from
$B^- \to D^{(*)0} \ell^- \bar{\nu}$ semileptonic decays (90\%) and rare $B$ decay
processes (10\%). 
About 30\% of the background has $K_L^0$ candidates in the KLM.

\begin{figure}
\begin{center}
 \includegraphics[width=.4\textwidth]{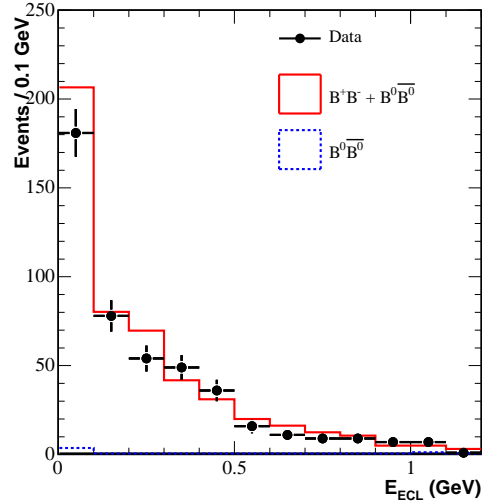}
\caption{$E_{\rm ECL}$ distribution for double tagged events, where one $B$ is fully
  reconstructed in the hadronic mode and the other $B$ is reconstructed as $B^{-} \rightarrow D^{*0}\ell^{-}\bar{\nu}$. 
  The dots indicate the data. The solid histogram is the background from 
  $B\overline{B}$ MC ($B^+B^- + B^0\overline{B}{}^0$), while the dashed one shows the contribution 
  from $B^0\overline{B}{}^0$ events.}
    \label{fig:controlsample}
\end{center}   
\end{figure}

In order to validate the $E_{\rm ECL}$ simulation, we use a control sample
of double tagged events, where the $B_{\rm tag}$ is fully reconstructed 
as described above and $B_{\rm sig}$ is reconstructed in the decay chain, 
$B^{-} \rightarrow D^{*0}\ell^{-}\bar{\nu}$ ($D^{*0}\rightarrow D^{0}\pi^{0}$),
followed by $D^0 \to K^- \pi^+$ or $K^- \pi^- \pi^+ \pi^+$
where $\ell$ is a muon or electron.
The sources affecting the $E_{\rm ECL}$ distribution in the control sample 
are similar to those in the signal MC simulation.
Fig.~\ref{fig:controlsample} shows the $E_{\rm ECL}$ distribution in the
control sample for data and the scaled MC simulation.
Their agreement demonstrates the validity of the $E_{\rm ECL}$ simulation 
in the signal MC.

\begin{figure}
\begin{center}
\includegraphics[width=.4\textwidth]{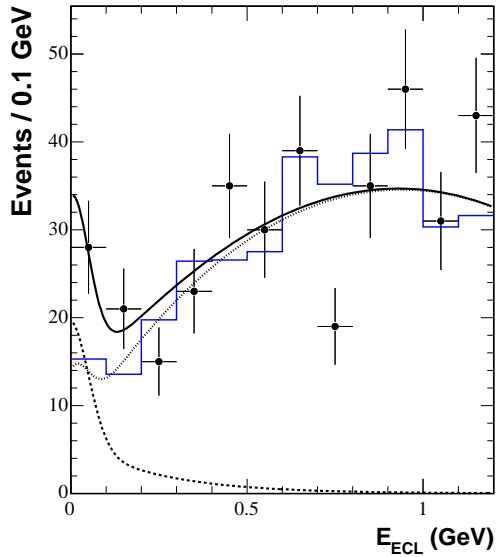} 
\caption{$E_{\rm ECL}$ distributions in the data after
	all selection criteria except the one on $E_{\rm ECL}$. 
	The data and background MC samples are represented by the points 
	and the solid histogram, respectively. 
	The solid curve shows the result of the fit with the sum of the signal (dashed) and 
	background (dotted) contributions.
}
    \label{ecl_opened}
\end{center}   
\end{figure}

After finalizing the signal selection criteria, the signal region is examined.
Fig.~\ref{ecl_opened} shows the $E_{\rm ECL}$ distribution obtained 
when all $\tau$ decay modes are combined.
One can see a significant excess of events in the $E_{\rm ECL}$ signal region
below $E_{\rm ECL}< 0.25$ GeV.
Table~\ref{tab:signal_yields} shows the number of events observed in the 
signal region ($N_{\rm obs}$) for each $\tau$ decay mode.
For the events in the signal region, we verify that the distributions of the 
event selection variables other than $E_{\rm ECL}$, such as $M_{\rm bc}$ and
$p_{\rm miss}$, are consistent with the sum of the signal and background
distributions expected from MC.
The excess remains after applying a $K_L^0$ veto requirement.

We obtain the final results by fitting the obtained $E_{\rm ECL}$ 
distributions to the sum of the expected signal and background shapes.
Probability density functions (PDFs) for the signal $f_{\rm s}(E_{\rm ECL})$ and
for the background $f_{\rm b}(E_{\rm ECL})$ are constructed for each $\tau$ decay
mode from the MC simulation.
The signal PDF is modeled as the sum of a Gaussian function, centered at
$E_{\rm ECL} = 0$, and an exponential function.
The background PDF, as determined from the MC simulation, is parameterized 
by  the sum of a Gaussian function and a second-order polynomial  function.
The Gaussian function in the background PDF addresses deviations 
from the second-order parameterization, which may arise from a peaking
component in the lower $E_{\rm ECL}$.
The PDFs are combined into an extended likelihood function,
\begin{equation}
{\cal L} = \frac{e^{-(n_{\rm s}+n_{\rm b})}}{N!}
\prod_{i=1}^{N}(n_{\rm s}f_{\rm s}(E_{i})+n_{\rm b}f_{\rm b}(E_{i})),
\end{equation}
where $E_{i}$ is the $E_{\rm ECL}$ in the $i$th event, $N$ is the total number 
of events in the data, and $n_{\rm s}$ and $n_{\rm b}$ are the signal yield and background 
yield to be determined by the fit 
to the whole $E_{\rm ECL}$ region ($0 < E_{\rm ECL} < 1.2$).
The results are listed in Table~\ref{tab:signal_yields}.
Table~\ref{tab:signal_yields} also gives the number of background events 
in the signal region deduced from the fit ($N_{\rm b}$), which is 
consistent with the expectation from the background MC simulation 
($N_{\rm sig}^{\rm MC}$).

The branching fractions are calculated as
${\cal B} = N_{\rm s}/(2\cdot\varepsilon\cdot N_{B^{+}B^{-}})$
where $N_{B^{+}B^{-}}$ is the number of $\Upsilon(4S)\rightarrow B^{+}B^{-}$ 
events, assuming $N_{B^{+}B^{-}} = N_{B^{0}\overline{B}^{0}}$.
The efficiency is defined as 
$\varepsilon = \varepsilon^{\rm tag}\times\varepsilon^{\rm sel}$,
where $\varepsilon^{\rm tag}$ is the tag reconstruction efficiency for events with 
$B^{-}\rightarrow\tau^{-}\bar{\nu}_{\tau}$ decays on the signal side, determined by MC
to be $0.136\pm 0.001({\rm stat})\%$,  and $\varepsilon^{\rm sel}$ is the event selection 
efficiency listed in Table~\ref{tab:signal_yields}, as 
determined by the ratio of the number of events surviving all the 
selection criteria including the $\tau$ decay branching fractions to the number of fully 
reconstructed $B^{\pm}$.
The branching fraction for each $\tau$ decay mode is consistent within errors.
To obtain the combined result for all $\tau$ decay modes, we
multiply the likelihood functions to produce the combined likelihood
(${\cal L}_{\rm com} = \prod_{j=1}^{5} {\cal L}_{j}$),
and constrain the five signal components by a single branching fraction.
The combined fit gives $17.2^{+5.3}_{-4.7}$ signal events in the signal 
region ($N_{\rm s}$) and $24.1^{+7.6}_{-6.6}$ in the entire region ($n_{\rm s}$).
The branching fraction is found to be $(1.79^{+0.56}_{-0.49}) \times 10^{-4}$.

\begin{table*}
  \renewcommand{\baselinestretch}{1.3}
    \begin{tabular}{cccccccccc} \hline\hline
&$N_{\rm side}^{\rm obs}$ &$N_{\rm side}^{\rm MC}$ &$N_{\rm sig}^{\rm MC}$ &$N_{\rm obs}$ &$N_{\rm s}$ &$N_{\rm b}$
&$\varepsilon^{\rm sel}(\%)$  &${\cal B}(10^{-4})$  &$\Sigma$\\\hline
$\mu^{-}\bar{\nu}_{\mu}\nu_{\tau}$ &$96$  &$94.2\pm 8.0$  &$9.4\pm 2.6$  &$13$ & $5.6^{+3.1}_{ -2.8}$  &$8.8^{ +1.1}_{  -1.1}$  &$3.64\pm 0.02$  &$2.57^{+1.38}_{-1.27}$ &$2.2\sigma$\\
$e^{-}\bar{\nu}_{e}\nu_{\tau}$     &$93$  &$89.6\pm 8.0$  &$8.6\pm 2.3$  &$12$ & $4.1^{+3.3}_{ -2.6}$  &$9.0^{ +1.1}_{  -1.1}$  &$4.57\pm 0.03$  &$1.50^{+1.20}_{-0.95}$ &$1.4\sigma$\\
$\pi^{-}\nu_{\tau}$                &$43$  &$41.3\pm 6.2$  &$4.7\pm 1.7$  &$9$  & $3.8^{+2.7}_{ -2.1}$  &$3.9^{ +0.8}_{  -0.8}$  &$4.87\pm 0.03$  &$1.30^{+0.89}_{-0.70}$ &$2.0\sigma$\\
$\pi^{-}\pi^{0}\nu_{\tau}$         &$21$  &$23.3\pm 4.7$  &$5.9\pm 1.9$  &$11$ & $5.4^{+3.9}_{ -3.3}$  &$5.4^{ +1.6}_{  -1.6}$  &$1.97\pm 0.02$  &$4.54^{+3.26}_{-2.74}$ &$1.5\sigma$\\
$\pi^{-}\pi^{+}\pi^{-}\nu_{\tau}$  &$21$  &$18.5\pm 4.1$  &$4.2\pm 1.6$  &$9$  & $3.0^{+3.5}_{ -2.5}$  &$4.8^{ +1.4}_{  -1.4}$  &$0.77\pm 0.02$  &$6.42^{+7.58}_{-5.42}$ &$1.0\sigma$\\
\hline										 										      
\hline
    \end{tabular}
    \caption{The number of observed events in data in the sideband region $(N_{\rm side}^{\rm obs})$,
      number of background MC events in the sideband region $(N_{\rm side}^{\rm MC})$ and the 
      signal region $(N_{\rm sig}^{\rm MC})$,
      number of observed events in data in the signal region $(N_{\rm obs})$, 
      number of signal $(N_{\rm s})$ and background $(N_{\rm b})$ in the signal region determined by the fit, 
      signal selection efficiencies $(\varepsilon^{\rm sel})$, 
      extracted branching fraction $({\cal B})$ for $B^{-}\rightarrow\tau^{-}\bar{\nu}_{\tau}$.
      The listed errors are statistical 
      only. The last column gives the significance of the signal including 
      the systematic uncertainty in the signal yield ($\Sigma$).}
   \label{tab:signal_yields}
\end{table*}

Systematic errors for the measured branching fraction are associated with 
the uncertainties in the  number of $B^{+}B^{-}$, signal yields and  
efficiencies.
The systematic error due to the uncertainty in $N_{B^{+}B^{-}}$ is $1$\%.
The uncertainty in the signal yields arises from uncertainties
in the signal and background shape, and is determined to be 
$^{+23}_{-26}$\%.
Here the uncertainty due to the signal shape uncertainty is determined 
by varying the signal PDF parameters by the amount of difference 
of each parameter between data and MC for the control sample of double tagged events.
To determine the background shape uncertainty, 
we vary the Gaussian constant of the background PDF by the branching fraction errors 
from PDG for the dominant peaking background sources (such as $B\rightarrow D^{(*)0}\ell\nu,
~ D^{0}\rightarrow \pi(K)\ell\nu$, etc.).
We then add in quadrature the variations for the signal and background
shapes.
We take a $10.5\%$ error as the systematic error associated with the tag reconstruction 
efficiency from the difference of yields between data and MC for the control sample.
This value includes the error in the branching fraction
${\cal B}(B^{-}\rightarrow D^{*0}\ell^{-}\bar{\nu})$, which we estimate from 
${\cal B}(B^{0}\rightarrow D^{*-}\ell^{+}\nu)$ in~\cite{Eidelman:2004wy}
and isospin symmetry.
The systematic error in the signal efficiencies depends on the $\tau$ decay
mode, and arises from the uncertainty in tracking efficiency ($1-3\%$), 
$\pi^{0}$ reconstruction efficiency ($3\%$), particle identification 
efficiency ($2-6\%$), branching fractions of $\tau$ decays 
($0.3-1.1\%$), and MC statistics ($0.6-2\%$).
These efficiency errors sum up to 5.6\% for the combined result after 
taking into account the correlations between the five $\tau$ decay modes~\cite{Lyons:1988rp}. 
The total fractional systematic uncertainty of the combined measurement is 
$^{+26}_{-28}\%$, 
and the branching fraction is
$$
{\cal B}(B^{-}\rightarrow\tau^{-}\bar{\nu}_{\tau}) 
= (1.79^{+0.56}_{-0.49}(\mbox{stat})^{+0.46}_{-0.51}(\mbox{syst}))\times 10^{-4}.
$$
The significance is $3.5\sigma$ when all $\tau$ decay modes are combined, 
where the significance is defined as 
$\Sigma = \sqrt{-2\ln({\cal L}_{0}/{\cal L}_{\rm max})}$,
where ${\cal L}_{\rm max}$ and ${\cal L}_{0}$ denote the maximum likelihood 
value and likelihood value obtained assuming zero signal events, respectively.
Here the likelihood function from the fit is convolved with a Gaussian
systematic error function in order to include the systematic uncertainty
in the signal yield.

In conclusion, we have found the first evidence of the purely leptonic decay
$B^{-}\rightarrow\tau^{-}\bar{\nu}_{\tau}$ from a data sample of $449\times 10^{6}$
$B\bar{B}$ pairs collected at the $\Upsilon(4S)$ resonance with the Belle 
experiment. 
The signal has a significance of $3.5$ standard deviations.
The measured branching fraction is
$(1.79^{+0.56}_{-0.49}(\mbox{stat})^{+0.46}_{-0.51}(\mbox{syst}))\times 10^{-4}$.
The result is consistent with the SM prediction within errors.
Using the measured branching fraction and known values of $G_F$, $m_B$, 
$m_{\tau}$~\cite{Eidelman:2004wy} and $\tau_B$~\cite{HFAG}, the product
of the $B$ meson decay constant $f_B$ and the magnitude of the 
CKM matrix element $|V_{ub}|$ 
is determined to be 
$f_{B} \cdot |V_{ub}| = (10.1^{+1.6}_{-1.4}(\mbox{stat})^{+1.3}_{-1.4}(\mbox{syst}))
\times 10^{-4}$ GeV.
Using the value of $|V_{ub}|$ from~\cite{HFAG}, we obtain 
$f_{B} = 0.229^{+0.036}_{-0.031}(\mbox{stat})^{+0.034}_{-0.037}(\mbox{syst})$ GeV,
the first direct determination of the $B$ meson decay constant.

We thank the KEKB group for excellent operation of the
accelerator, the KEK cryogenics group for efficient solenoid
operations, and the KEK computer group and
the NII for valuable computing and Super-SINET network
support.  We acknowledge support from MEXT and JSPS (Japan);
ARC and DEST (Australia); NSFC and KIP of CAS (contract No.~10575109 
and IHEP-U-503, China); DST (India); the BK21 program of MOEHRD, and the
CHEP SRC and BR (grant No. R01-2005-000-10089-0) programs of
KOSEF (Korea); KBN (contract No.~2P03B 01324, Poland); MIST
(Russia); ARRS (Slovenia);  SNSF (Switzerland); NSC and MOE
(Taiwan); and DOE (USA).




\begin{thebibliography}{99}
\bibitem{conjugate}
Charge conjugate states are implied throughout the paper.
\end{thebibliography}
\end{document}